# Grant-Free Non-Orthogonal Multiple Access:
# A Key Enabler for 6G-IoT

Rana Abbas, The University of Sydney, Tao Huang, James Cook University, Muhammad Basit Shahab, The University of Newcastle, Mahyar Shirvanimoghaddam, The University of Sydney, Yonghui Li, The University of Sydney, and Branka Vucetic, The University of Sydney

*The proliferating number of devices with short payloads as well as low power budget has already driven researchers away from classical grant-based access schemes that are notorious for their large signalling overhead as well as power-consuming retransmissions. Instead, light-weight random access protocols have been re-investigated and their throughput has been improved in orders of magnitude with sophisticated yet still low-complex transceiver algorithms. In fact, grant-free access has been identified as a key medium access control technique for providing massive connectivity in machine type communications in cellular networks. In this paper, we show that grant-free access combined with non-orthogonal transmission schemes is a promising solution for 6G Internet of Things (IoT). We present novel and promising results for deep learning (DL)-based techniques for joint user detection and decoding. Then, we propose a multi-layered model for GF-NOMA for power-efficient communications. We also discuss resource allocation issues to enable the co-existence of GF-NOMA with other orthogonal or even grant-based schemes. Finally, we conclude with proposed research directions for medium access towards enabling 6G-IoT.*

## Introduction

The fifth generation (5G) of cellular networks is to provide three main services: enhanced Mobile Broadband (eMBB), massive Machine Type Communications (mMTC). and ultra-Reliable Low Latency Communications (uRLLC). The latter two services are expected to enable a wide range of transformative applications ranging from vehicular communications (V2X), to Industry 4.0, smart grids, remote health monitoring etc. On the other hand, research on the sixth generation (6G), albeit still a decade away, is envisioning a whole new range of applications that fall under *ubiquitous wireless intelligence* [1] which will enormously increase telepresence as well as autonomous vehicles in transport and logistics. This requires addressing the short comings of 5G in the technical KPIs including (1) reducing the end-to-end latency by 1/10 (from 1 ms down to 0.1 ms), (2) supporting massive connectivity for 10,000x traffic and (iii) increasing the power efficiency by 10 folds [1]. In this paper, we focus on achieving this KPI on the multiple access (MAC) layer. Thus, it is first appropriate to review the proposed and adopted medium access protocols for 5G.

### Medium Access Protocols in 5G

Medium access in 5G systems will involve a wider spectrum (up to 60 GHz) and will allow for some non-orthogonal techniques (yet to be determined). However, it will mostly still be OFDMA-based. However, there is large consensus in the literature that this almost fully orthogonal multiple access approach will not suffice to support the massive access of devices, especially as we slowly move towards cell-less architectures [2].

Moreover, we know that 5G access will be mostly grant-based and random access will only be used for acquiring grant requests, i.e. establishing a connection. However, to reduce the signalling overhead, we want machine type devices (MTDs) to be delivering their payloads through random access. That is because grant-based communications in cellular networks suffer from collision rates over the random access channel that are as high as 10%, for less than 10 active users. Moreover, the signalling overhead involved in establishing a link is about 30-50% of the payload size, for messages less than 200 bits long [3]. In terms of latency, the grant-based access procedure in LTE-A takes around 5-8 ms, in the best-case scenario, excluding the time required for data transmission and assuming error-free and collision-free signalling [3]. Thus, grant-based access fails at meeting many KPIs when massive connectivity is required for short packet transmissions.

Considering this, two types of grant-free UL transmission are specified in Rel-15, namely Type 1 and Type 2 [4]. For Type 1, no L1 signalling is required to kick-start transmission on a configured UL resource. In contrast, Type 2 essentially follows the principle of LTE semi-persistent scheduling (SPS) operation suitable for periodic or semi-periodic traffic. We have strong to believe that efforts on grant-free access will continue for the next decade as experts at the first 6G Wireless Summit [1] concluded that novel grant-free access methods are essential for massive connectivity. Although 5G will support some grant-free access methods, the predicted massive number of low-power and low-complexity devices will need to be serviced with a much higher spectral efficiency in 6G. The other obstacle is the energy consumption of battery-powered devices that are expected to run advanced signal-processing algorithms on large amounts of data. The latter is mostly driven by artificial intelligence (AI)-based solutions which will be at the core of 6G.

The standardised lower power solutions for mMTC such as extended coverage GSM (EC-GSM), LTE for machine-type communication (LTE-M), and narrow band IoT (NB-IoT) that operate on top of existing cellular networks are all still grant-based. Moreover, the random access protocols governing the link establishment process is conceptually ALOHA-like. The same applies for the contention-based medium access protocols in WiFi, HaLow, SigFox, and LoRaWAN.

Thus, there has been no "revolution" in medium access control in 5G. However, both industry and academia agree that fundamentally new paradigms are much needed for enabling beyond 5G (B5G) systems [1]. In this paper, we show why such paradigms should involve Grant-Free Non-Orthogonal Access (GF-NOMA) techniques. GF-NOMA is designed to embrace interference by a fine cross-layer integration of the physical (PHY) and medium access control (MAC) functionalities Our discussions focus mostly on mMTC services, though can be extended to URLLC [3].

## Grant-Free Non-Orthogonal Multiple Access

NOMA allows multiple users to share the different resources, e.g., time, frequency, and space, either through power domain multiplexing or code domain multiplexing. Thus, unlike orthogonal multiple access, overloading is possible at the expense of increased processing complexity at the receiver. Moreover, in the finite block length regime, orthogonal multiple access (OMA) is strictly suboptimal in terms of spectrum efficiency from an information-theoretic point of view.

Grant-free NOMA (GF-NOMA) is fundamentally different than the heavily investigated coordinated NOMA. The main issue in GF-NOMA is that the set of active users and sometimes their respective channel conditions are not known at the receiver. These parameters also change randomly from one slot to the next. Moreover, channel estimation, synchronization and multi-user detection (MUD) should be in one shot for optimal performance. These unknowns limit and often prohibit the pre-assignment of pilots/preambles, power, codebooks and many more forms of classical resource allocation.

Existing GFA Schemes can be categorized into three main categories described below. We refer interested readers to [5] for an in-depth survey of existing schemes in each category.

- *Signature Sequence-based*: Signature sequences can be spreading, scrambling and/or interleaving sequences with the common property of orthogonality or near-orthogonality to allow multi-user detection. This is almost always needed when channel state estimation is required at the receiver.
- *Compute-and-forward-based:* Users encode their messages with two concatenated channel codes; one for error correction, and one for user detection. The inner code achieves MUD at the receiver by allowing unique mapping between the transmitted set of messages and the received

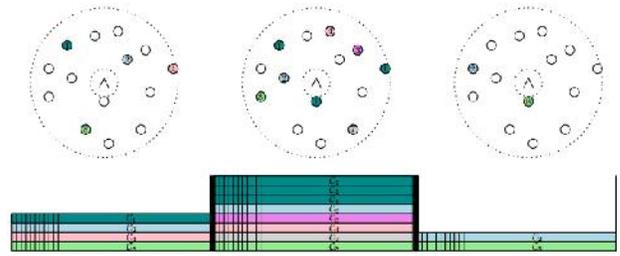

**Figure 1** *Schematic Diagram of Grant-Free Non-Orthogonal Multiple Access showing the sporadic nature of traffic. The number of users in each frame is not generally known to the receiver. The different colours correspond to packet transmissions with different signature sequences, chosen randomly by the devices from a common pool. Packets that are transmitted with the same signature sequence are said to have collided and fail to be recovered.*

  sum. It also reduces the complexity by bypassing the need to decode every message separately.
- *Artificial Intelligence-based:* These algorithms look for patterns in data for making nearly optimal decisions with practical levels of complexity. It can also be a useful design tool for end-to-end system performance, especially when the input space is large.

### Joint User Detection and Decoding

In GF-NOMA, the receiver's task is to first detect how many users were active (transmitting) in the received block. Then, the receiver will estimate their channels. This is often referred to as blind estimation to emphasize the receiver's little knowledge of the state of the network. Provided correct user detection, the receiver continues to perform demodulation and decoding. Decoding can either be joint using maximum likelihood decoding, belief propagation or the same as that of point to point by leveraging successive-interference cancellation (SIC) techniques.

### Common Signature Sequences

One essential design parameter in GF-NOMA is the contention transmit unit (CTU). A contention transmit unit can be multi-dimensional covering time slots, frequency bands, codebooks and signature sequences. For many years, these signature sequences were chosen to be orthogonal with excellent auto and cross-correlations. Ideally, auto-correlations of orthogonal sequences should result in a delta function whose amplitude is used for channel estimation, and cross-correlations amongst distinct sequences should result in an all-zero sequence. The most common example of such sequences is the *Zadoff-Chu (ZC) Sequences* used in LTE.

A ZC sequence is based on a primer sequence of complex values characterized by its index u length q. In existing cellular networks, ZC sequences of length 839 are used and the total number of available sequences is less than 64 (with some reserved for contention-free transmissions by high-priority applications).

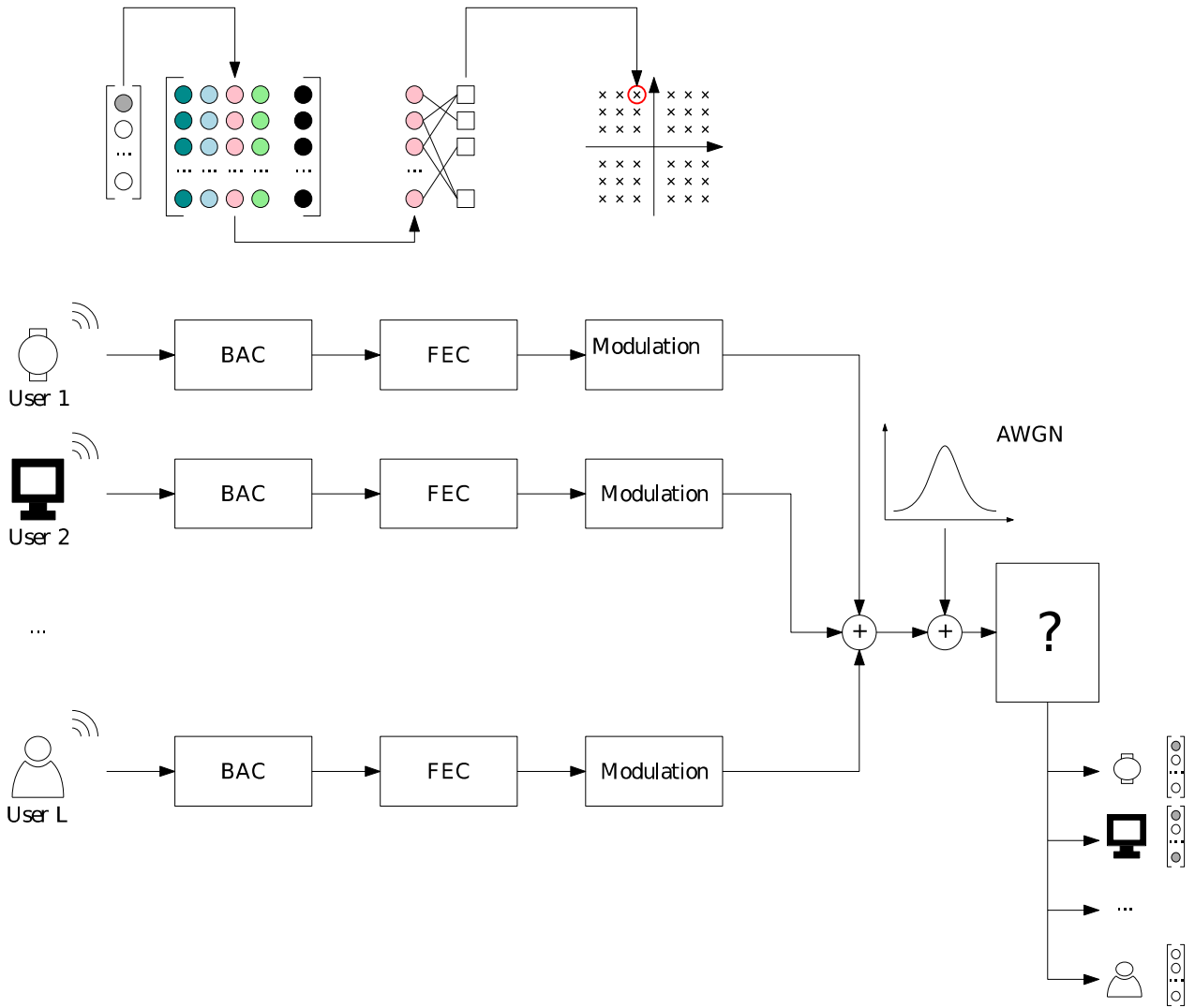

**Figure 2** *Proposed System Model for GF-NOMA*

In general, the main drawback of ZC sequences is their focus on orthogonality which restricts the set of sequences that can be made available for a given length. With sophisticated algorithms, one can ensure the same reliability in user detection while allowing a higher level of correlation for a larger pool of sequences. One good example of such codes is the second-order *Reed-Muller (RM) Code,* where a sequence of length $2^m$ can generate up to $2^{m(r+2)}$ sequences [5].

For the case where the transmitters' received powers are the same, the receiver's main task is reduced to detecting the number of transmitting users and recovering the list of messages that were sent. From information theory, we know that there exists code that can provide unique mappings between sets of messages and their sum. The parity check matrix of BCH codes are examples of such codes. If users choose their sequences from the columns of a binary *BCH code* parity check matrix of error correction capability T, the sum of their transmitted codewords can be decoded with a small error probability provided that (1) the sum of their sequences can be correctly decoded at the receiver side, (2) no more than T users have transmitted simultaneously and (3) all users choose distinct sequences. Existing low complex detection procedures can be implemented with a computational cost of only $O(kT^2)$ additions and multiplications in $GF(2k)$. In the next section, we compare these detection methods with a DL-based detector and show that we can achieve even better spectral efficiency while maintaining low complexity.

## Deep Learning for GF-NOMA

Recent research continues to confirm the powerful capabilities of AI in enhancing transceiver designs in wireless communications [6]. For NOMA systems, AI has been applied to several of its prohibitively complex problems such as (1) channel state information acquisition, (2) resource allocation, (3) clustering or user pairing, (5) complex joint decoding, etc. This is especially useful in massive NOMA, as the complexity of these processes grows exponentially with the number of users.

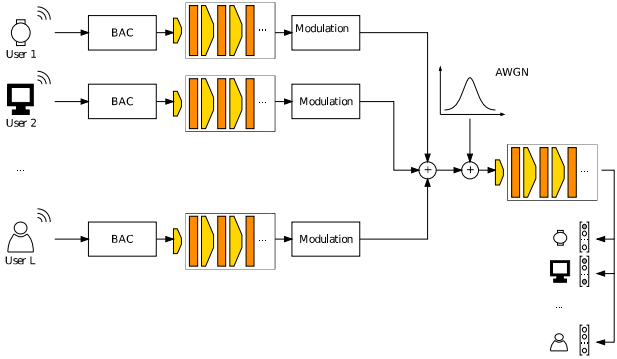

**Figure 3** *Fully Connected Neural Network Architecture*

However, there is very little that has been done in the space of GF-NOMA [5]. The setup that follows can hopefully fill in some of the gap in the literature by demonstrating the capabilities of deep learning in a simple yet generic framework. The results can also be a much-needed benchmark for more sophisticated models and transmissions schemes to come.

## Deep-Learning Network Architecture

We consider a system with N users, out of which a small yet random set of users are active at one given time. The active users transmit simultaneously over the same time-frequency block using the same codebook. Users map their messages of size k to the columns of a parity check matrix of a BCH code ($2^k$ -1, kT), where T is the error correction capability. This mapping represents a channel code of rate 1/T. For simplicity, we assume that they choose unique codewords [7], as repeated codewords will cancel each other out and violate the one to one mapping between the sum and the individual messages transmitted. To further improve the reliability of the system, the users can further encode their codewords with a powerful error correction code of rate R to combat the noisy channel. We consider three different detection techniques that represent different trade-offs between complexity and performance: the MLD (high complexity, high reliability), the Berlekamp Massey algorithm (BMA) (low complexity, low reliability), and a fully connected neural network (low complexity, high reliability).

For the DL-based detector, we adopt the structure of the auto-encoder [6]. As shown in Figure 3, there is an encoder at each user, and a decoder at the receiver. Note that the encoder used at each user is the same. We used 4 hidden layers, where each layer consists of 2048 neurons. The activation function used for each hidden layer in the encoder is ReLU and the activation function used for its output layer is a Sigmoid function. The output of the encoder is BPSK-modulated and amplified. For the decoder, we also used 4 hidden layers and each layer consists of 2048 neurons. The activation function for each hidden layer is ReLU. The output layer of the decoder contains $2^k$-1 nodes, and the loss function used is a binary cross-entropy function. The learning rate is 0.0001 and the Adam optimizer was used. The total number of training data set is $2 \times 10^5$ and the number of validation data set is 104. The number of test data is $10^7$. The training is performed at $E_b/N_0$ = 8 dB and across different cases with different number of active users (less or equal T). The number of epochs is 1000 and each mini-batch is $10^5$. The goal is to train the encoder and decoder simultaneously so that the decoder can output the correct collection of messages that have been transmitted by the users.

## DL-based Detector Performance

We define the block error rate as the probability that a transmitted message is not recovered at the receiver. Our results demonstrated in Figure 4 a remarkable improvement in performance between the deep learning approach and the BMA case. When comparing against MLD, we observe that there is still a margin for improvement. Moreover, as the network needs to be trained for a specific signal to noise ratio (SNR), we can see error floors. In practice, it is desirable to attain the best performance at each SNR. However, that would require different trained models to be implemented at the receiver for each desired SNR. Otherwise, the model will always perform poorly over SNRs that it has not experienced before. Training over multiple SNRs is a potential solution; however, in this case, we found that it confused the decoder and let to a poorer performance. On the other hand, the cost of implementing multiple trained models at the receiver side is not clear and needs further investigation.

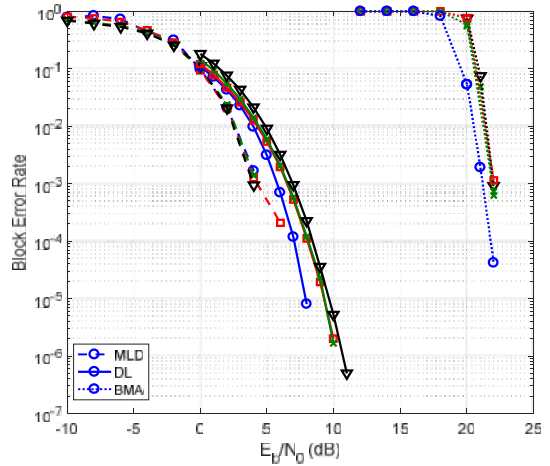

**Figure 4** *The block error rate is computed for a (63,39) BCH code (T = 4) and R = ½. For MLD and BMA, this represents the case where the columns of BCH matrix are further encoded with a convolutional code of memory order 7.*

An interesting case is depicted in Figure 5, where the code rate R is set to 1. For the case of BMA and MLD, we take this as the uncoded case where the columns of the parity check matrix are modulated and transmitted directly. However, for the DL case, the system can have the same complexity for both encoding and decoding as the case before and still add a forward error correction code of rate 1. Thus, we see, for the same complexity we can have a coding gain that allows the DL-based detector to outperform MLD in this case.

Finally, in Figure 6, we demonstrate the flexibility of the DL-based detector in targeting different performance gains depending on the application requirements. We retrain the model based on a weighted loss function. We set the loss function to be the weighted sum of loss functions obtained by training the model at 8 dB and 12 dB. We weight the losses for the case of 3 users and 4 users, by 10 and 20, respectively. We also weight the overall losses at 12 dB by 4. The reason behind these choices is to improve the error floors exhibited in Figure 4. We note here that the optimal choice of loss function remains an open problem [6]. Results show that by improving the case of 3 and 4 users, the total loss is improved. Thus, we conclude that DL is flexible in terms of its training target. However, the downside is that we cannot predict the model performance without testing the trained neural network.

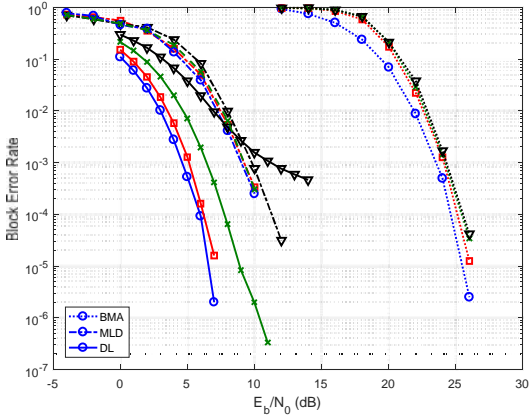

**Figure 5** *The block error rate is computed for a (63,39) BCH code (T = 4) and R = 1. For MLD and BMA, this represents the case where the columns of the BCH matrix are not further encoded.*

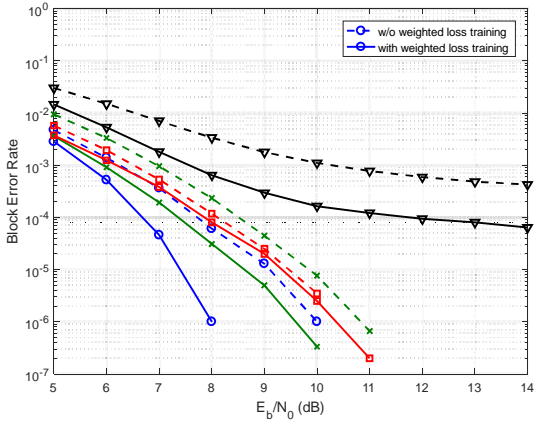

**Figure 6** *Results for DL-based detector with different loss functions. Same setup as Figure 4.*

## Power Allocation

The previous setup assumes that all participating devices tune to the same received power level at the. This is possible only when the is channel reciprocal, e.g. time-division multiplexing. That is because the user needs to have accurate channel state information without explicitly sending requests to the receiver prior to data transmission. When this is possible, enforcing the same received power has many advantages including reducing the length of the pilot sequences as well as reducing the error probabilities in detection [8]. However, MTDs are at risk of frequent outages, depending on its location and transmit power constraint. Alternatively, we can say that some MTDs are denied service more than others. Moreover, to increase service availability, the network needs to provide more flexibility. This can be done by allowing devices with lower channel state information to tune to received powers that suit their state.

To support multiple received powers, we propose a layered approach [7], where layer carries out the same encoding/decoding process, only with a different power. Layers can be detected using pilot sequences, and this assignment can be broadcasted as frequently as needed to MTDs. The number of layers should not be too large as this is a main requirement for NOMA to work well.

In Figure 7, we consider the case of two layers only. With only two layers, the pilot sequences can be a reasonable size while ensuring negligibly low detection error rates. We set a threshold on the maximum received power allowed ($P_{max}$) and optimise the power allocations to minimise the maximum error probability. This multi-layer approach clearly supported more devices while maintaining the same complexity, as layers can be decoded separately. It also shows a power-saving, as users can tune their powers to lower levels and still access the channel reliably. This is especially important for MTDs that may not be able to tune their power to the selected received power level.

In a more practical setting, with variable arriving packets in each slot, there are multiple power allocation schemes that can be devised without explicitly identifying active users. We propose two simple solutions below for the special case where the packet arrival distribution and rate are known by the AP.

- *Channel-Based Power Allocation Scheme*: For a maximum transmit power, the users can choose the layer that requires least transmit power based on their channel gain.
- *Random Power Allocation Scheme*: Alternatively, the users can pick at random will properly designed probabilities, amongst the layers which satisfy their transmit power constraint,

One important consequence of the transmit power constraint, is that there will always be a non-zero probability that some users will not be able to satisfy any of the required received power levels. In that case, these users cannot transmit. Alternatively, as in cellular networks, a mixed loop power control scheme can be employed where users that cannot meet the required received power level transmit with their maximum transmit power. In this scenario, as the power of each layer needs to be accurately estimated at the receiver side, the AP can allocate some pilot sequences for these users. However, note that if two or more of these users choose the pilot sequence, the AP will fail to recover these users and the linearity of the code is in question. Interested readers are referred to [8] for this model.

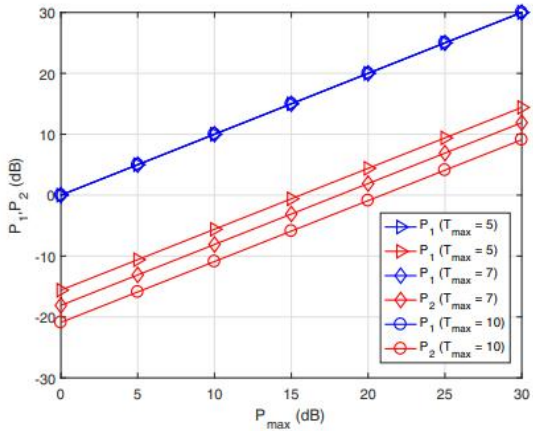

**Figure 7** Optimal Power Allocation for a Multi-Layered Architecture of GF-NOMA through Mixed Integer Linear Programming [7]

## Resource Allocation for GF-NOMA

In order to enable any form of grant-free access, the radio resource for transmission should be defined before any GF transmission starts and be known to both the users and the receiver. The predefined resource for GF transmission can be termed as a GF resource unit (GFRU) or a contention transmission unit (CTU, [9]). The GFRUs/CTUs comprise of a time-frequency physical resource combined with a set of pilots for channel estimation and/or user activity detection, and a set of multiple access (MA) signatures for robust signal transmission and interference whitening. The MA signatures set may include at least one of the following; codebook/codeword, sequence, interleaver and/or mapping pattern, demodulation reference signal, power-dimension, spatial-dimension, preamble, etc. [9]. Once the structure of GFRUs is designed, the size and location of time/frequency resources, as well as the pilot/signature patterns associated with them should be pre-defined and known to both the devices and the receiver. One example of a CTU/GFRU is shown in Figure 8, which comprises of time-frequency resource, codebooks, and pilots.

Once the basic resource is defined, the resource allocation problem is to study how to allocate these resources to different users. In this context, the selection of a GFRU by users can be predefined or predetermined [10]. Furthermore, in order to enable autonomous GFA, the users can also randomly and independently choose a GFRU to transmit their data [10]. Once these resources are selected by the users through any of the mentioned mechanism, the users can then perform link adaptation i.e., selecting suitable modulation and coding scheme (MCS), transmit power, etc., and transmit their data. Normally, link adaptation is associated with the channel information, which is not exactly known to the receiver and users in GFA. To resolve this, one solution is to consider the channel to be reciprocal in each direction, as was done in the previous section. Hence, users can estimate their channel to the BS using the periodically received pilot/reference signals in downlink, and correspondingly adjust their uplink transmission parameters to facilitate data recovery at the BS. Another solution is to divide the radio resources into MA blocks/groups (MABs), where resources in a MAB correspond to specific transmit parameter settings i.e., transmission block sizes, MCSs, transmit power, etc. Configurations of the MABs can be broadcasted by the BS. During the data transmission phase, any active user first selects a MAB followed by using the associated parameters for data transmission. At the receiver, when the BS receives a particular MAB, it exactly knows what transmission parameters are used by the user, and therefore performs data recovery efficiently by considering these parameters.

One important point to be taken into consideration during resource allocation is the coexistence of human type communication (HTC) and MTC devices. Contrary to MTC, HTC mostly revolves around high data rates, where dedicated/orthogonal resource allocation with centralized scheduling-based access (grant-based) may be needed. Hence, the coexistence of contention-based NOMA and scheduled access is supported by various studies in academia and industry. For practical purposes, only a portion of uplink bandwidth can be configured as contention/grant-free regions, while the other portion can be used for regular scheduled uplink data transmissions. The size and number of access regions are dependent on many factors e.g., expected number of MTC/HTC devices and/or applications/services, etc.

One major challenge that exacerbates the resource allocation problem is the diverse quality of service (QoS) of the IoT/MTC devices. In such scenarios, a uniform resource allocation GFA policy may not be a feasible solution to the issue. In this context, resource pool partitioning is an efficient solution, where multiple NOMA sub-regions/partitions can be defined within the NOMA resource pool [4]. Furthermore, the users can be clustered/grouped based on their QoS requirements; all users with similar QoS can be clustered together. Once this is done, different resource pool partitions can be dedicated to different user clusters/groups to satisfy their QoS requirements. This is similar to the concept of network slicing, which aims to provide the flexibility of resource allocation to different use cases of IoT framework. The objective is to allow a physical mobile network operator to partition its network resources to allow for very different users to multiplex over a single physical infrastructure. The most commonly cited example in 5G discussions is the sharing of a given physical network to simultaneously run mMTC, eMBB, and URLLC. NOMA with network slicing has been under focus recently. In [11], the vital challenges of resource management pertaining to network slicing using the NOMA-based scheme are highlighted. In this context, efficient solutions for resource management in network slicing for NOMA-based scheme are provided. Moreover, a slice-based virtual resource scheduling scheme with NOMA to enhance the QoS of the system is proposed in [12].

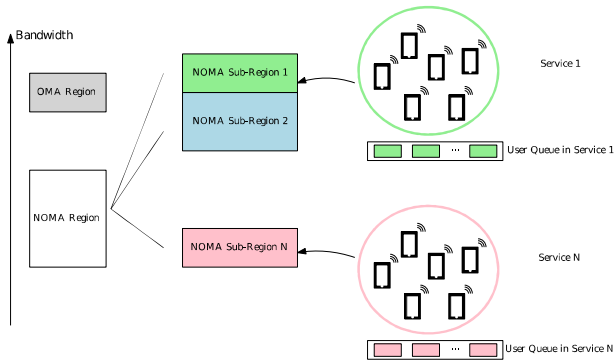

**Figure 8** *Resource Allocation for a hybrid access scheme: consisting of OMA and NOMA techniques*

A generalized illustration of resource allocation by considering the aforementioned points is depicted in Figure 8. The overall system bandwidth or time-frequency resources are divided into NOMA and OMA regions. Moreover, the NOMA regions are further divided into multiple sub-regions/slices/partitions to provide grant free access using GFRUs/CTUs. These subregions are then allocated to various user clusters, where each cluster consists of the mMTC devices with similar QoS requirements. To add more flexibility for resource allocation and considering a practical scenario where the number of users in each service type can change, the OMA sub-regions can also be changed dynamically as per needs. In this way, a more efficient network management to provide GFA to a diverse QoS based IoT framework can be achieved.

## Conclusion

In conclusion, we demonstrated in this paper the capabilities of deep learning techniques to improve detection and decoding in GF-NOMA settings with reasonable complexity. Results show that deep learning can approach MLD, the optimal decoder, with much reasonable complexity. More importantly, deep learning is a powerful tool to design code parameters for end-to-end target performances. Although optimal training parameters are far from obvious for most cases, they present incredible flexibility in meeting different targets as demonstrated in this paper. As grant free access is much needed to support massive connectivity in 6G, our results herein can serve as an excellent benchmark for future proposals in this field. In the remaining part of the paper, we discuss resource allocation issues for GF-NOMA and propose a multi-layered approach to enable more users in a power efficient manner.

It is important to note that this paper considered one-shot transmissions only. That is, the presented block error rates are the probabilities of failure of the first transmission attempt. However, in practice, automatic repeat request (ARQ) and/or hybrid ARQ (HARQ) protocols will be implemented to enhance the reliability of data transmission. Thus, one future research direction is optimising retransmission protocols using AI. However, HARQ protocols can be spectrally inefficient as rates are reduced dramatically with every retransmission attempt.

Alternatively, 6G calls for tighter integrations between the data frame structure and the error correction codes as well as robust coding and modulation schemes to the limited channel state information available at both transmitters and receivers [1]. This was also demonstrated in this paper by the sub-optimal performance of DL in SNR regions that were not included in the training. To this end, we believe the time has come for self-adaptive coding techniques to be implemented at the physical layer whose coding structure depends very little on the channel state. So far, such self-adaptive codes, also dubbed rateless codes, such as RaptorQ, have been limited to the network layer (as standardised in DVB-S). However, more modern and more powerful rateless codes can be useful for both link adaptation as well as load adaptation. Recent results [13] show that they can exhibit excellent performance in comparison to traditional channel codes, implying that the self-adaptive property can be achieved with minimal performance loss. Applications of such codes to medium access control are few [14] yet no doubt growing mostly due to advances in information theory aiming at deriving their fundamental limits and, thus, proving analytically their superiority in this space [5].

Rana Abbas [M] (rana.abbas@sydney.edu.au) received the M.E. in 2013 and the Ph.D. degree in 2018, both in electrical engineering, from The University of Sydney. She is currently a researcher at the Centre of IoT and Telecommunications, at The University of Sydney. Her research interests include channel coding, random access, machine type communications and IIoT. She is the recipient of the Australian Postgraduate Awards Scholarship and the Norman I. Price scholarship from the Centre of Excellence in Telecommunications, School of Electrical and Information Engineering, The University of Sydney. She is also the winner of the Best Paper Award at IEEE PIMRC, 2018.

Tao Huang (tao.huang1@jcu.edu.au) received the Ph.D. degree in Electrical Engineering from The University of New South Wales, in 2014, and stayed there as post-doctoral research fellow till 2015. He was an Endeavour Australia Cheung Kong Research Fellow supported by the Commonwealth Government of Australia in 2014. In 2018, he joined the College of Science and Engineering, James Cook University, where he is currently a lecturer in Electronic Systems and IoT Engineering. He has co-authored a Best Paper Award at 2011 IEEE Wireless Communications and Networking Conference. He is a co-inventor of one patent on MIMO systems.

Muhammad Basit Shahab (basit.shahab@newcastle.edu.au) received his B.S. in Electrical Engineering from University of Engineering and Technology (UET) Lahore, Pakistan. His M.S. in Electrical Engineering was from University of Management and Technology (UMT) Lahore, Pakistan. He finished his Ph.D. in February 2019 from the Department of IT Convergence Engineering, Kumoh National Institute of Technology (KIT), South Korea. Currently, he is working as a postdoctoral research fellow at the School of Electrical Engineering and Computing, The University of Newcastle (UoN), Australia. His main research areas are radio access technologies (RATs) for future wireless communications, non-orthogonal multiple access (NOMA), cooperative communication, half/full-duplex relaying, and M2M/D2D communications.

Mahyar Shirvanimoghaddam [M] (mahyar.shm@sydney.edu.au) received the B.Sc. degree (1st Class Honours) from the University of Tehran, Iran, in September 2008, the M.Sc. degree (1st Class Honours) from Sharif University of Technology, Iran, in October 2010, and the Ph.D. degree from The University of Sydney, Australia, in January 2015, all in electrical engineering. He then held a postdoctoral research position at the School of Electrical Engineering and Computing at the University of Newcastle, Australia. He is currently an academic fellow at the School of Electrical and Information Engineering, The University of Sydney, Australia. His general research interests include channel coding techniques, multiple access techniques, and machine-to-machine communications.

Yonghui Li [F] (yonghui.li@sydney.edu.au) received his Ph.D. degree in November 2002 from Beijing University of Aeronautics and Astronautics. Since 2003, he has been with the Centre of Excellence in Telecommunications, the University of Sydney, Australia. He is now a professor in the School of Electrical and Information Engineering, University of Sydney. He was the recipient of the Australian Queen Elizabeth II Fellowship in 2008 and the Australian Future Fellowship in 2012. His current research interests are in the area of wireless communications, with a particular focus on MIMO, millimeter wave communications, machine to machine communications, coding techniques and cooperative communications.

Branka Vucetic [F] (branka.vucetic@sydney.edu.au) is an ARC Laureate Fellow and professor of telecommunications, and Director of the Centre of Excellence in Telecommunications at the University of Sydney. During her career, she has held research and academic positions in Yugoslavia, Australia, UK and China. Her research interests include coding, communication theory and signal processing and their applications in wireless networks and industrial Internet of Things. She has co-authored four books and more than 400 papers in telecommunications journals and conference proceedings. She is a Fellow of the Australian Academy of Technological Sciences and Engineering and a Fellow of the IEEE.